\documentclass[preprint,aps,epsfig,tighten]{revtex4}
\usepackage{graphicx,epsfig,axodraw,rotating}
\begin{document}


\title{Meson formfactor scheme for the chiral Lagrangian approach to 
$J/\psi$ breakup cross sections motivated by a relativistic quark model}

\author{D.~B.~Blaschke}
\affiliation{
Institute for Theoretical Physics, University of Wroc{\l}aw,
50-204 Wroc{\l}aw, Poland}
\affiliation{
Bogoliubov Laboratory for Theoretical Physics, JINR Dubna,
141980 Dubna, Russia}
\author{H.~Grigorian}
\affiliation{Laboratory of Information Technologies, JINR Dubna, 141980 Dubna,
Russia}
\author{Yu.~L.~Kalinovsky}
\affiliation{Laboratory of Information Technologies, JINR Dubna, 141980 Dubna,
Russia}
\date{\today}

\begin{abstract}
We suggest a new scheme for the introduction of formfactors for
the $SU(4)$ chiral meson Lagrangian approach to
the  $J/\psi$ breakup cross sections by pion and rho meson impact.
This mesonic formfactor scheme respects the fact that on the quark level of
description the contact and the meson exchange diagrams are constructed by
so-called box and triangle diagrams which contain a different number of
vertex functions for the quark-meson coupling.
We present a model calculation for Gaussian vertex functions within the meson
formfactor scheme and compare the results with those of the usual global
formfactor model.
We calibrate the new meson formfactor model with results for the pion impact
processes from a relativistic quark model calculation by Ivanov et al. and
present predictions for the $\rho$-meson induced processes.
We provide a fit formula for the resulting energy-dependent cross sections.
\vspace{5mm}
\noindent
\end{abstract}
\maketitle

\section{Introduction}
The $J/\psi$ meson plays a key role in the experimental search for the
quark-gluon plasma (QGP) in heavy-ion collision experiments where an
anomalous suppression of its production cross section relative to the
Drell-Yan continuum as a function of the centrality of the collision has
been found by the CERN-NA50 collaboration \cite{Abreu:2000ni}.
An effect like this has been predicted to signal QGP formation
\cite{Matsui:1986dk}
as a consequence of the screening of color charges in a plasma in close
analogy to the Mott effect (metal-insulator transition) in dense electronic
systems \cite{Redmer:1997}.
However, a necessary condition to explain $J/\psi$ suppression in the static
screening model is that a sufficiently large fraction of $c\bar c$ pairs after
their creation have to traverse regions of QGP where the temperature
(resp. parton density) has to exceed the Mott temperature
$T^{\rm Mott}_{{\rm J}/\psi}\sim 1.2 - 1.3~ T_c$
\cite{Karsch:1987pv,Ropke:1988bx} for a
sufficiently long time interval $\tau>\tau_{\rm f}$,  where
$T_c\sim 190$ MeV is the critical phase transition temperature and
$\tau_{\rm f}\sim 0.3 $ fm/c is the $J/\psi$ formation time.
Within an alternative scenario \cite{Ropke:1988zz},
$J/\psi$ suppression does not
require temperatures well above the deconfinement one but can occur already
at $T_c$ due to impact collisions by quarks from the thermal medium.
An important ingredient for this scenario is the lowering of the reaction
threshold for string-flip processes which lead to open-charm meson formation
and thus to $J/\psi$ suppression.
This process has an analogue in the hadronic world, where e.g.
J/$\psi + \pi \rightarrow D^* + \bar D + h.c.$ could occur provided the
reaction threshold of $\Delta E \sim 640$ MeV can be overcome by pion impact.
It has been shown \cite{Blaschke:2000er,Burau:2001pn}
that this process and its in-medium
modification can play a key role in the understanding of anomalous $J/\psi$
suppression as a deconfinement signal.
Since at the deconfinement transition the $D$- mesons enter the continuum of
unbound (but strongly correlated) quark- antiquark states (Mott- effect),
their spectral function is broadened and the relevant threshold for charmonium
breakup is effectively lowered so that the reaction rate for
the process gets critically enhanced \cite{Blaschke:2003ji}.
Thus a process which is almost negligible in the vacuum may give rise to
additional (and therefore anomalous) $J/\psi$ suppression when
conditions of the chiral/ deconfinement transition and $D$- meson Mott effect
are reached in a heavy-ion collision.
The dissociation of the $J/\psi$ itself still needs impact to overcome the
threshold which is still present but dramatically reduced.

For this alternative scenario \cite{Blaschke:2000er} to work the $J/\psi$
breakup cross section by meson impact is required and its
dependence on the masses of the final state $D$- mesons has to be determined.
After the first calculations of this quantity within a nonrelativistic
potential model (NPM) \cite{Martins:1995hd} and its systematic  improvement
\cite{Wong:1999zb}, the relativistic quark model (RQM) calculation suggested
in \cite{Blaschke:2000zm} could recently be completed for $J/\psi$
dissociation processes by pion impact \cite{Ivanov:2003ge}.
Both approaches result in energy dependent cross sections with a steep rise
to a maximum of the order of $1$ mb close to the threshold followed by a fast
drop.
This agreement in shape and magnitude is a nontrivial result since the NPM
contains quark and gluon exchange diagrams only at first Born order whereas
the RQM includes meson exchange diagrams between the colliding mesons, which
correspond to ladder-type resummation of gluon exchanges.

Shortly after the publication of the NPM results, a chiral Lagrangian (CL)
approach to the problem of $J/\psi$ dissociation has been suggested
\cite{Matinyan:1998cb} which resulted in qualitatively different predictions
for the magnitude and the energy dependence of the dissociation cross sections:
a monotonously rising behaviour which reached the level of $1$ mb only about
400 MeV above the threshold.
In this work, however, important processes have not been considered and
subsequent, more systematic developments of the CL approach
\cite{Lin:1999ad,Haglin:1999xs,Oh:2000qr,Lin:2000jp,Haglin:2000ar,Bourque:2004av}
have revealed the importance of $D^*$ meson exchange and contact diagrams.
Their inclusion lead to a steep rise of the cross section at the threshold to
a level of several tens of millibarns followed by a continuous rise with
increasing energy.
It has been admitted, however, that this monotonously rising behaviour of the
cross section is an artifact of the treatment of mesons as pointlike
particles in the CL approach.
Phenomenological formfactors have therefore been introduced in order to take
into account the finite size of meson-meson vertices due to the composite
nature of the mesons.
Unfortunately, the results of the formfactor-improved CL models (FCLM) appear
to be strongly dependent on the choice of those formfactors for the
meson-meson vertices
\cite{Lin:1999ad,Lin:2000jp,Haglin:2000ar,Ivanov:2001th,Oh:2002vg,Bourque:2008ta}.
This is a basic flaw of these approaches which could only be
overcome when a more fundamental approach, e.g., from a quark model, can
determine these input quantities of the chiral Lagrangian approach.
There have been attempts to reduce the uncertainties of the
FCLM by constraining the choice of the formfactor using a comparison
with results of the NPM approach which makes use of meson wave
functions \cite{Ivanov:2001th,Oh:2002vg}.

In the present work we suggest a formfactor model which takes into account
the different sizes of mesons in the collision by meson-dependent range
parameters and accounts for the fact that in a RQM the contact diagrams
are represented by a quark loop containing four mesonic wave functions
(box diagram) which suppress the transition amplitude at
high momentum transfer, while meson exchange diagrams are represented by
the product of triangle diagrams containing six meson wave functions.
We calibrate the parameters of this meson formfactor model by comparison
with a RQM calculation of the $J/\psi$ dissociation by pion impact
\cite{Ivanov:2003ge}.
On the basis of this newly developed meson formfactor CL (MFCL) model we make
a prediction for the rho-meson impact dissociation processes of the $J/\psi$
meson.
We suggest that this MFCL model can be applied for the calculation
of the in-medium modification of the  $J/\psi$ breakup due to the
Mott-effect for mesonic states at the deconfinement/chiral restoration
transition according to the quantum kinetic approach suggested in
\cite{Blaschke:2000er,Burau:2001pn} and provide an explanation of
the anomalous $J/\psi$ suppression effect observed in heavy-ion
collisions at the CERN-SPS \cite{Abreu:2000ni}.
The MFCL approach developed here can be applied also to $J/\psi$ dissociation
by nucleon impact which is of central importance for the quark matter
diagnostics under dense nuclear matter conditions as, e.g., in the planned
CBM experiment at FAIR Darmstadt or at NICA Dubna.

\section{$J/\psi$ breakup cross sections for $\pi$ and $\rho$ meson impact
from the chiral Lagrangian approach}
\label{sec:cross}

\begin{figure}[htb]
\epsfig{figure=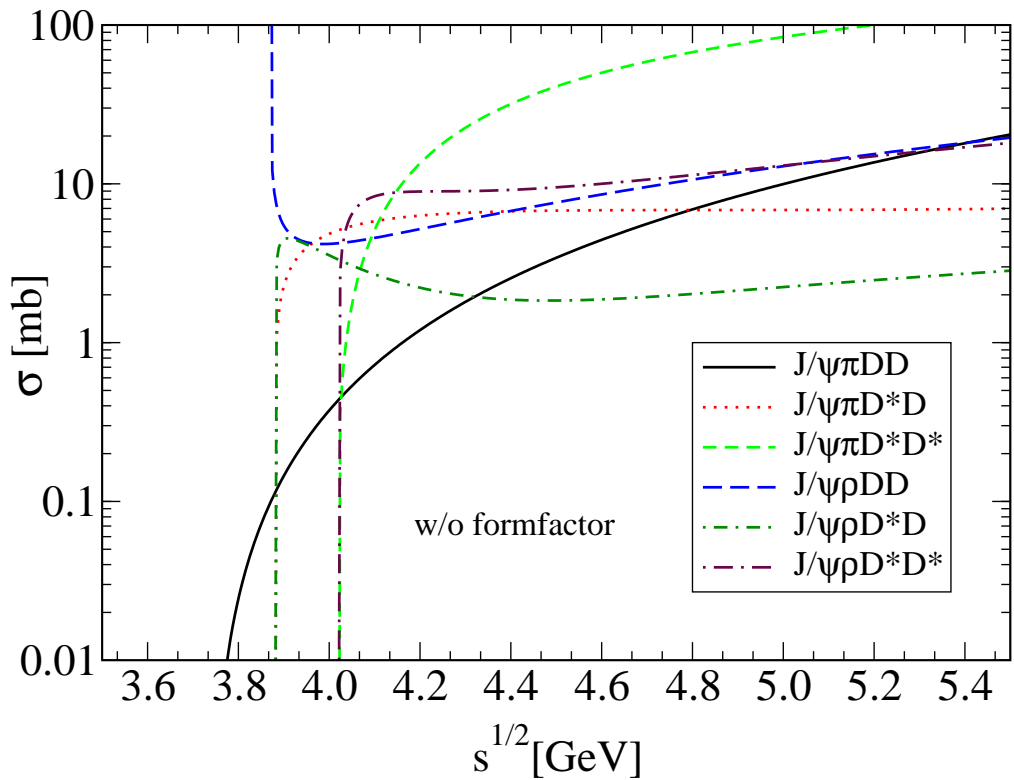,height=0.95\textwidth,angle=-90}
\caption{(Color online) $J/\psi$ break-up cross sections  by pion and rho-meson
impact in the chiral Lagrangian model without formfactors.
\label{fig1}}
\end{figure}

The effective Lagrangian approach we employ in the present study is
developed in Ref.~\cite{Oh:2000qr} based on the minimal $SU(4)$ Yang-Mills
Lagrangian including anomalous parity interactions which are connected to the
gauged Wess-Zumino action, namely $\pi+J/\psi\to D+\bar D$,
$\pi+{\rm J/}\psi\to D^*+\bar D^*$ and $\rho + J/\psi\to D^*+\bar D$.
The processes we discuss in the following on this basis are thus
\begin{eqnarray}
\label{ps1}
&& J/\psi + \pi\rightarrow
D+\bar{D}, D^* + \bar{D}, D + \bar{D}^*, D^*+\bar{D^*}  \\
&& J/\psi + \rho \rightarrow
D+\bar{D}, D^* + \bar{D}, D + \bar{D}^*, D^*+\bar{D^*}.
\label{ps2}
\end{eqnarray}
The corresponding cross sections  after averaging (summing) over initial
(final) spins and including isospin factors can be represented in a form
given by

\begin{eqnarray}
\label{cross}
\displaystyle \frac{d\sigma}{dt}
= \frac{1}{ 64\pi s p^2_{cm} I_s I_i }
M_{\lambda_k\ldots \lambda_l}M_{*\lambda_k'\ldots \lambda_l'}
N^{\lambda_k \lambda_k'}_k(p_k)\ldots N^{\lambda_l \lambda_l'}_l(p_l)~,
\end{eqnarray}
where the  $s,t,u$ are the standard Mandelstam variables,
\begin{eqnarray}
p^2_{cm} = \frac{[s-(m_1+m_2)^2][s-(m_1-m_2)^2]}{4s}
\end{eqnarray}
is the squared momentum of initial state mesons in the center-of-momentum
frame.
In Eq. (\ref{cross}) we have introduced the projectors
\begin{eqnarray}
\label{project}
N^{\lambda \lambda'}_k(q)
= \left( g^{\lambda\lambda'} - \frac{q^{\lambda}q^{\lambda'}}{m_k^2}\right)
= \varepsilon^\lambda(q)\cdot \varepsilon^{*\lambda'}(q)
\end{eqnarray}
corresponding to each involved vector particle of species $k$ and their 
relationship to the polarisation vectors $\varepsilon^\lambda(q)$ employed, 
e.g., in Ref.~\cite{Oh:2000qr}. 
This notation makes explicit that each amplitude contributing to the 
differential cross section (\ref{cross}) is manifestly gauge invariant
$p^{\lambda_k}M_{\lambda_k\ldots \lambda_l}=0$. 
The factors $I_s$ and $I_i$ result from averaging over initial spins and
isospins, respectively. Their evaluation amounts to  $I_sI_i = 3/2$
for the processes in Eq.~(\ref{ps1}) and to $I_sI_i = 9/2$ for those in
Eq.~(\ref{ps2}).
The amplitudes
$M_{\lambda_k\ldots \lambda_l}=
\sum_{j=1}^s M_{\lambda_k\ldots \lambda_l}^{(j)}$
of the corresponding processes are collected in tables given in the Appendix.
The index $j$ specifies the diagram which contributes to a process with
given initial and final states.
In the tables we use the $Q$ for the momentum of the exchanged $D$ and
$D^*$ mesons. For $t-$channel exchange we have $Q=p_1-p_3$ while for
$u-$channel exchange $Q=p_2-p_3$ holds.
The results for the energy dependent cross sections of the $J/\psi$ breakup
processes by pion and rho-meson impact are shown in Fig. \ref{fig1}.
The unknown coupling constants of the chiral Lagrangian approach are fixed
according to the scheme given in Ref.~\cite{Oh:2000qr} which is based on the
$SU(4)$ symmetry relations, the vector dominance model and the observed decay
$D^* \longrightarrow D \pi$. The resulting values of \cite{Oh:2000qr} are
given in the Table of Appendix \ref{constants}.

As it has been discussed in the Introduction, the monotonous rise of the
cross section with increasing energy has to be considered as an unphysical
artifact of the neglect of the finite size of the mesons due to their
quark substructure.
In the following Section we suggest a new scheme for introducing
hadronic formfactors which is based on the insight from the NPM (RQM)
that they originate from the meson wavefunctions (Bethe-Salpeter amplitudes)
involved in the representation of the meson vertices by quark exchange
(quark loop) diagrams.

\section{Hadronic Formfactors}

The chiral Lagrangian aproach for $J/\psi$ breakup by light meson impact
makes the assumption that mesons and meson-meson interaction vertices are
pointlike (four-momentum independent) objects. This neglect of the finite
extension of mesons as quark-antiquark bound states has dramatic
consequences: it leads to a monotonously rising behaviour of the cross
sections for the corresponding processes, see  Fig. \ref{fig1}.

This result, however, cannot be correct away from the reaction threshold
where the tails of the mesonic wave functions determine the high-energy
behaviour of the quark exchange (in the nonrelativistic formulation of
\cite{Martins:1995hd,Wong:1999zb}) or quark loop (in the relativistic
formulation \cite{Blaschke:2000zm,Ivanov:2003ge}) diagrams describing the
microscopic processes underlying the $J/\psi$ breakup by meson impact.
Since the mesonic wave functions
describing quark-antiquark bound states have a finite extension in
coordinate- and momentum space, the $J/\psi$ breakup cross section is
expected to be decreasing function for high c.m. energies of the collision.
Such a behaviour has been obtained within NPM and RQM approaches to
meson-meson interactions
\cite{Martins:1995hd,Wong:1999zb,Blaschke:2000zm,Ivanov:2003ge}.
In order to model such a behaviour within
chiral meson Lagrangian approaches one uses formfactors at the
interaction vertices \cite{Lin:1999ad,Haglin:2000ar,Ivanov:2001th,Oh:2002vg}
which can be calibrated using quark model results for available processes.
On this basis the cross sections for otherwise unknown processes can be
predicted.
When the amplitudes of all subprocesses are separately gauge invariant
(see the discussion of Eqs. (\ref{cross}) and (\ref{project}) in 
Sect.~\ref{sec:cross}), the procedure of their rescaling with different 
three-momentum dependent formfactors does not violate the gauge invariance of 
the chiral Lagrangian approach.
However, the violation of the transversality requirement by assuming different 
formfactors for processes (which are not separately transveral as motivated 
for example by the relativistic quark model) should not be mismatched with a 
possible violation of the conservation laws by the considered processes. 
Due to the implementation of phenomenological formfactors our approach 
qualifies as an effective one for which conservation laws are fulfilled
by the construction of the amplitudes for interaction vertices.

\subsection{Global formfactor ansatz}

\begin{figure}[htb]
\begin{tabular}{cc}
\epsfig{figure=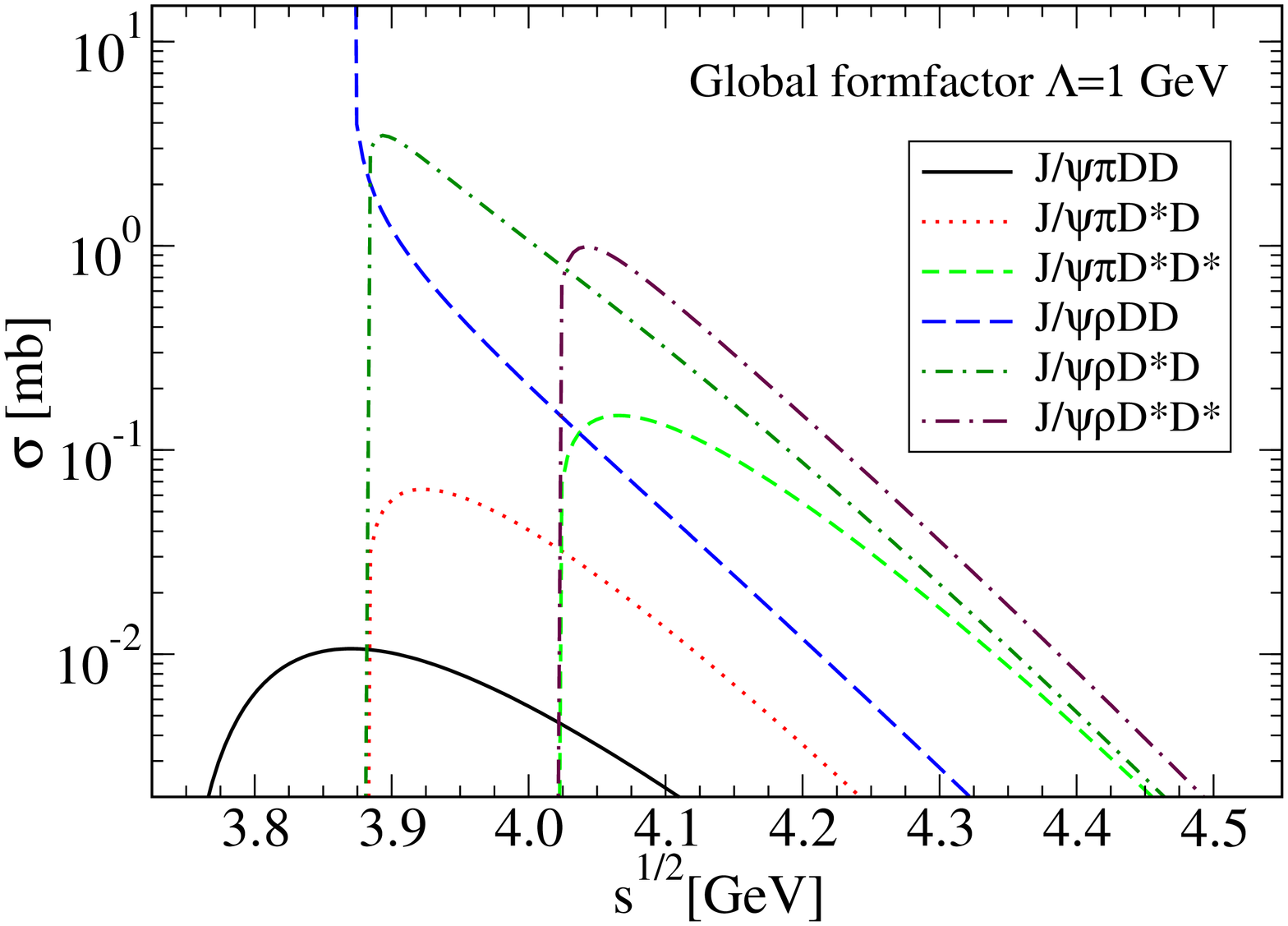,height=0.45\textwidth,angle=-90}&
\epsfig{figure=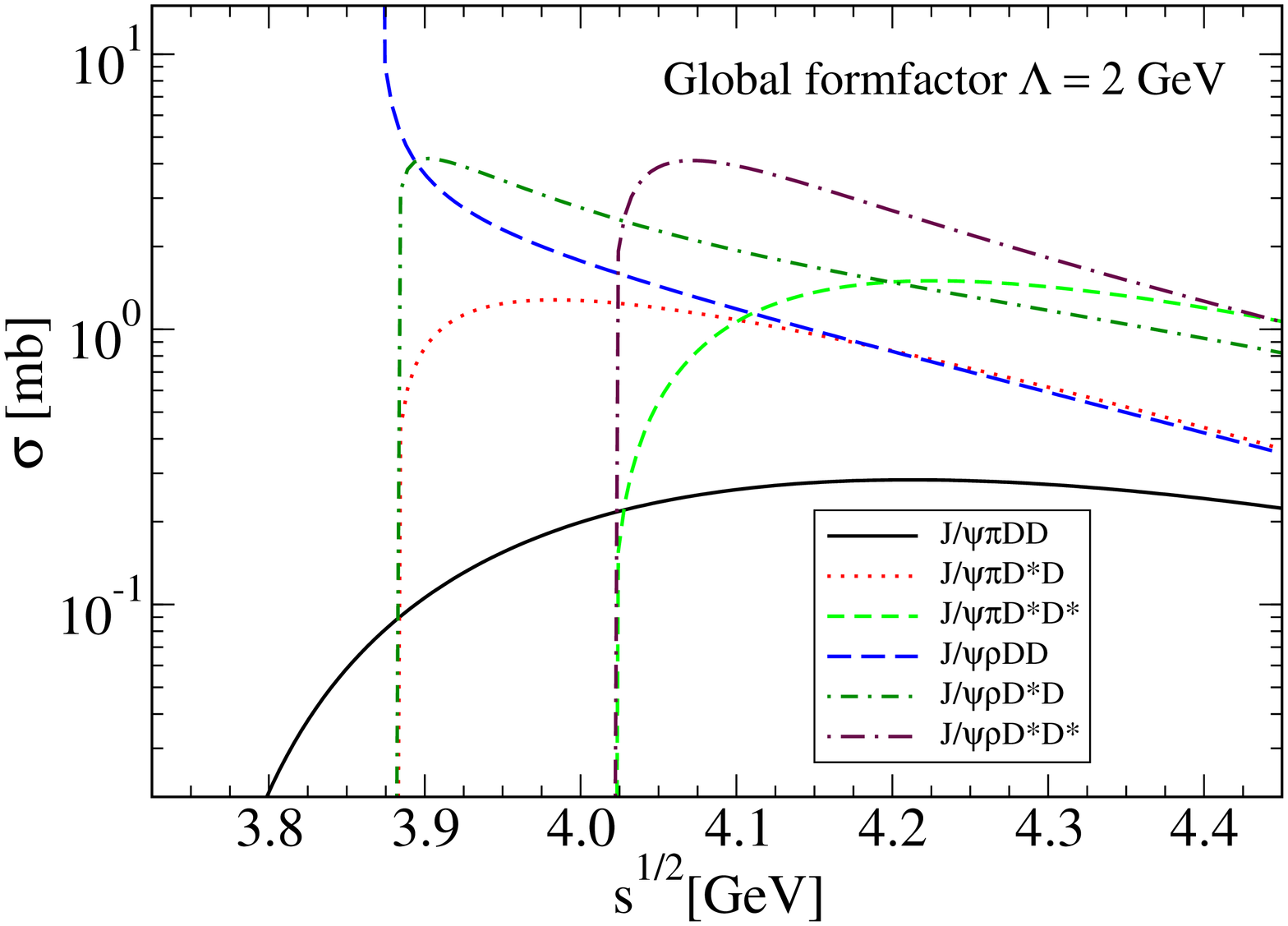,height=0.45\textwidth,angle=-90}
\end{tabular}
\caption{(Color online) $J/\psi$ break-up cross sections in the chiral
Lagrangian model with Gaussian global formfactors and range parameters
$\Lambda=2$ GeV (left panel) and $\Lambda=1$ GeV (right panel).
\label{fig2}}
\end{figure}

The simplest ansatz for a hadronic formfactor disregards the specifics of
different mesonic species such as the different radii of their wave functions.
Following the definitions of Ref. \cite{Lin:1999ad}, the formfactor of all the
four-meson vertices given in the Appendix \ref{amplitudes}, i.e. those of the
contact diagrams as well as those
of the meson exchange diagrams is given in the same form.
It is represented as a product of the three-meson
vertices and intermediate meson propagation
\begin{equation}
\label{f3f3} F_4({\bf q}^2)=\left[F_3({\bf q}^2)\right]^{2}~~.
\end{equation}
In this formula, ${\bf q}^2$ is given by the average value of
the squared three-momentum transfers in the $t$ and $u$ channels
\begin{equation}
{\bf q}^2=\frac{1}{2}\left[({\bf p_1}-{\bf p_3})^2+ ({\bf p_2}-{\bf
p_3})^2\right]_{\rm c.m.}= p^2_{i,{\rm c.m.}}+p^2_{f,{\rm c.m.}}~.
\end{equation}
Here, $p_{i,{\rm c.m.}}$ and $p_{f,{\rm c.m.}}$ can be represented by
using the Mandelstam variables $s$
\begin{eqnarray}
p^2_{i,{\rm c.m.}}&=&\frac{1}{4 s}
\Big( s - (m_1 + m_2)^2 \Big) \Big( s - (m_1 - m_2)^2 \Big),\nonumber\\
p^2_{f,{\rm c.m.}}&=&\frac{1}{4 s}
\Big( s - (m_3 + m_4)^2 \Big) \Big( s - (m_3 - m_4)^2 \Big).
\label{pcm}
\end{eqnarray}
For the three-meson vertices, we use formfactors with a momentum dependence
in the Gaussian ($G$) form
\begin{equation}
F_3({\bf q}^2)=\exp(-{{\bf q}^2/\Lambda^2})~,
\end{equation}
motivated by the behavior of a mesonic bound state wave function.
The results for the  $J/\psi$ breakup cross sections by light meson impact
with this global formfactor model are shown in Fig.~\ref{fig2}.

\subsection{Meson formfactor ansatz}
\label{ssec:mesonform}

In order to take into account the quark substructure of meson-meson vertices
we suggest here a simple ansatz which respects the different sizes of the
interacting mesons and the different quark diagram representations in terms of
quark box and quark triangle diagrams.
The triangle diagram is of third order in the wave functions so that the meson
exchange diagrams are suppressed at large momentum transfer by six wave
functions while the box diagram appears already at fourth order thus being
less suppressed than suggested by the ansatz (\ref{f3f3}) of Ref.
\cite{Lin:1999ad}.

For the contact terms (four-meson vertices) we introduce
according to the scheme $g_{J/\psi \pi D^* D} \longrightarrow
g_{J/\psi \pi D^* D} \times F_c(s)\,$
the contact formfactors $F_c(s)$ in the following form

\begin{eqnarray}
F_c(s) = \exp\Bigg\{
-\Bigg[p^2_{i,{\rm c.m.}}
\bigg( \frac{1}{\Lambda_1^2}+ \frac{1}{\Lambda_4^2} \bigg) + p^2_{f,{\rm c.m.}}
\bigg( \frac{1}{\Lambda_3^2} + \frac{1}{\Lambda_4^2} \bigg)
\Bigg]
\Bigg\}\;
\end{eqnarray}

The formfactors $F_u(s,t)$ for the $u-$ channel meson exchange terms are
introduced according to the example
$g_{J/\psi D^* D^*} \times g_{D^* D \pi}
\longrightarrow g_{J/\psi D^* D^*} \times g_{D^* D \pi} \times F_{u}(s,t)\,$
with the form

\begin{eqnarray}
F_{u}(s,t) = \exp\Bigg\{
-\Bigg[p^2_{i,{\rm c.m.}}
\bigg( \frac{1}{\Lambda_1^2}+ \frac{1}{\Lambda_4^2} \bigg) + p^2_{f,{\rm c.m.}}
\bigg( \frac{1}{\Lambda_3^2} + \frac{1}{\Lambda_4^2} \bigg)
+ \frac{2{\bf q}(s,t)^2}{\Lambda_4^2}
\Bigg]
\Bigg\}\; .
\end{eqnarray}

Analogously,  the formfactor $F_t(s)$ for the $t-$ channel meson exchange
diagrams is introduced according to the example
 $g_{J/\psi D D} \times g_{D^* D \pi}
\longrightarrow g_{J/\psi D D} \times g_{D^* D \pi} \times F_{t}(s,u)\,$
with the formfactor

\begin{eqnarray}
F_{t}(s,u) =\exp\Bigg\{
-\Bigg[p^2_{i,{\rm c.m.}}
\bigg( \frac{1}{\Lambda_1^2}+ \frac{1}{\Lambda_2^2} \bigg) + p^2_{f,{\rm c.m.}}
\bigg( \frac{1}{\Lambda_3^2} + \frac{1}{\Lambda_4^2} \bigg)
+ \frac{2{\bf q}(s,u)^2}{\Lambda_3^2}
\Bigg]
\Bigg\}\;
\end{eqnarray}

and is related to the $F_u(s,t)$ by exchanging $u \leftrightarrow t$ and
$\Lambda_3 \leftrightarrow \Lambda_4$.
Here the ${\bf q}^2(s,t)$ can be also rewritten using the Mandelstam
variables $s,t$ and $u$ as

\begin{equation}
\label{qt}
{\bf q}^2(s,t)=
\bigg( \Big( m_1^2 - m_2^2 \Big) - \Big( m_3^2 - m_4^2 \Big)\bigg)^2/4 s
-  t ;
\end{equation}

for the $t-$channel and
\begin{equation}\label{qu}
{\bf q}^2(s,u)=
\bigg( \Big( m_1^2 - m_2^2 \Big) + \Big( m_3^2 - m_4^2 \Big) \bigg)^2/4 s
-  u ;
\end{equation}

for the $u-$channel meson exchange processes.
The dependences of the formfactors on $p^2_{i,{\rm c.m.}}$,
$p^2_{f,{\rm c.m.}}$ on the transferred
momentum ${\bf q}$ are the same as in the global formfactor case, see
Eq.(\ref{pcm}).
Here we use the phenomenological range parameters
$\Lambda_i=\alpha \Lambda^0_i$ of the
meson-quark-antiquark vertices which shall resemble the ranges of the
corresponding meson wave functions in momentum space and thus be closely
related to the meson masses $m_i$, see Table \ref{tab:mass}.
The physical meaning of such an approach is to take
into account that for high energies the cross-section of given process is
suppressed due to the lack of time for quark exchange between interacting
mesons as well as due to the reduction of the overlap of meson wave functions.
The ansatz of the effective range according to the rule
$\bigg(1/{\Lambda_1^2}+ 1/{\Lambda_2^2} \bigg)$ means that in the case of
different meson sizes the amount of suppression is dominated by heavier,
i.e. the smaller meson.
Such an ansatz is in accord with the phenomenological Povh-H\"ufner law for
total hadron-hadron cross sections   \cite{Povh:1990ad}.
For later use, we introduce additionally a common scale factor $\alpha$ to be
used in fixing the formfactor by comparison with the RQM approach of 
Ref.~\cite{Ivanov:2003ge}.

\begin{table}
 \begin{ruledtabular}
 \begin{tabular}{cccccc}
  {state $i$} & $J/\psi~ $ & $D^*~ $ & $D~ $ & $\varrho~ $ & $\pi~ $ \\
  \hline \hline
  {$m_i [{\rm GeV}]$} & 3.1 & 2.01 & 1.87 & 0.77 & 0.14 \\
  {$\Lambda^0_i [{\rm GeV}]$} & 3.1 & 2.0 & 1.9 & 0.8 & 0.6 
 \end{tabular}
 \end{ruledtabular}
\caption{(Color online) Meson masses and range parameters corresponding to the
quark-antiquark-meson vertices as used in the meson formfactor ansatz
of Subsection~\ref{ssec:mesonform}.
\label{tab:mass}
}
\end{table}

\begin{figure}[htb]
\begin{tabular}{cc}
\epsfig{figure=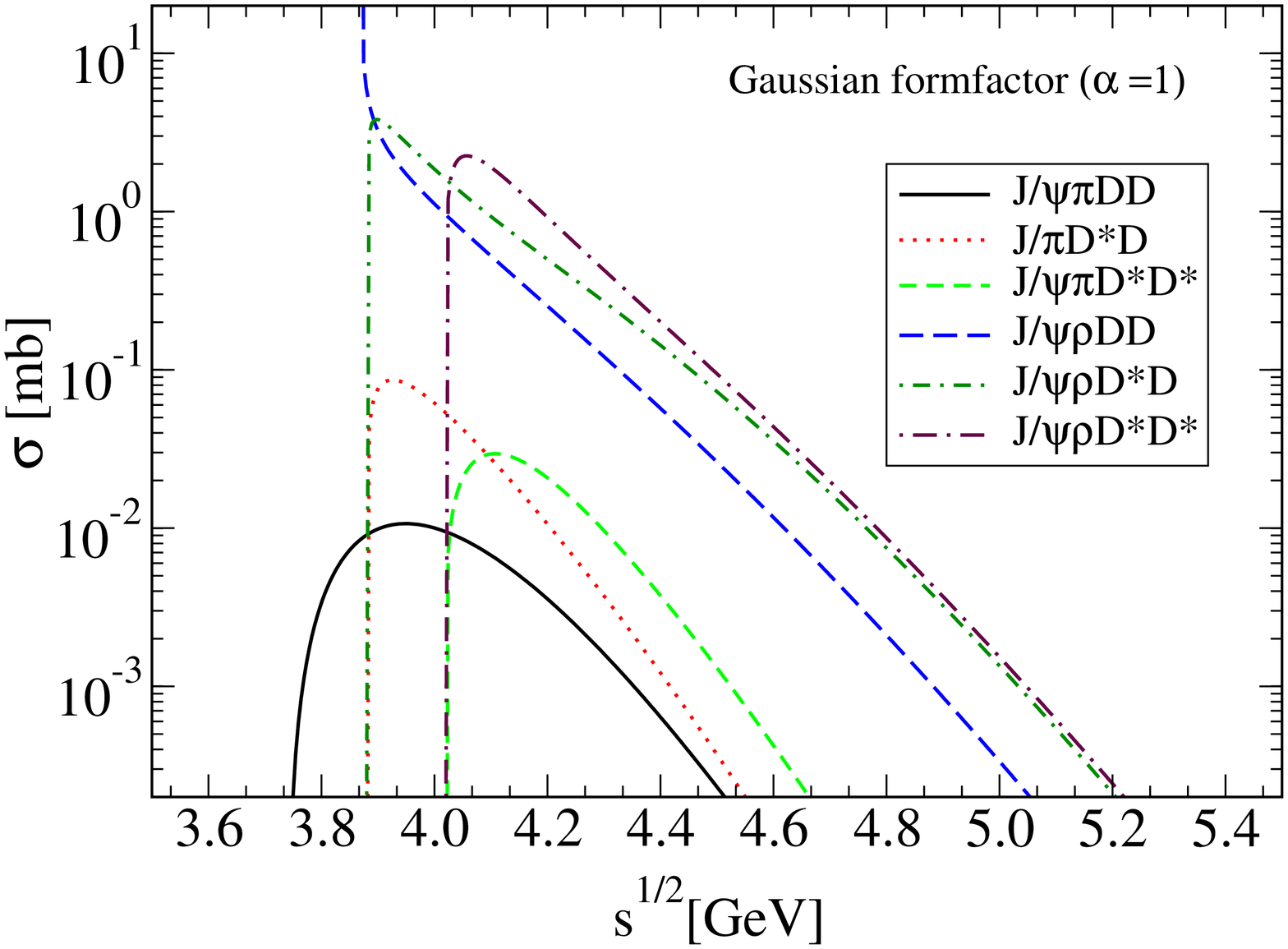,height=0.45\textwidth,angle=-90}&
\epsfig{figure=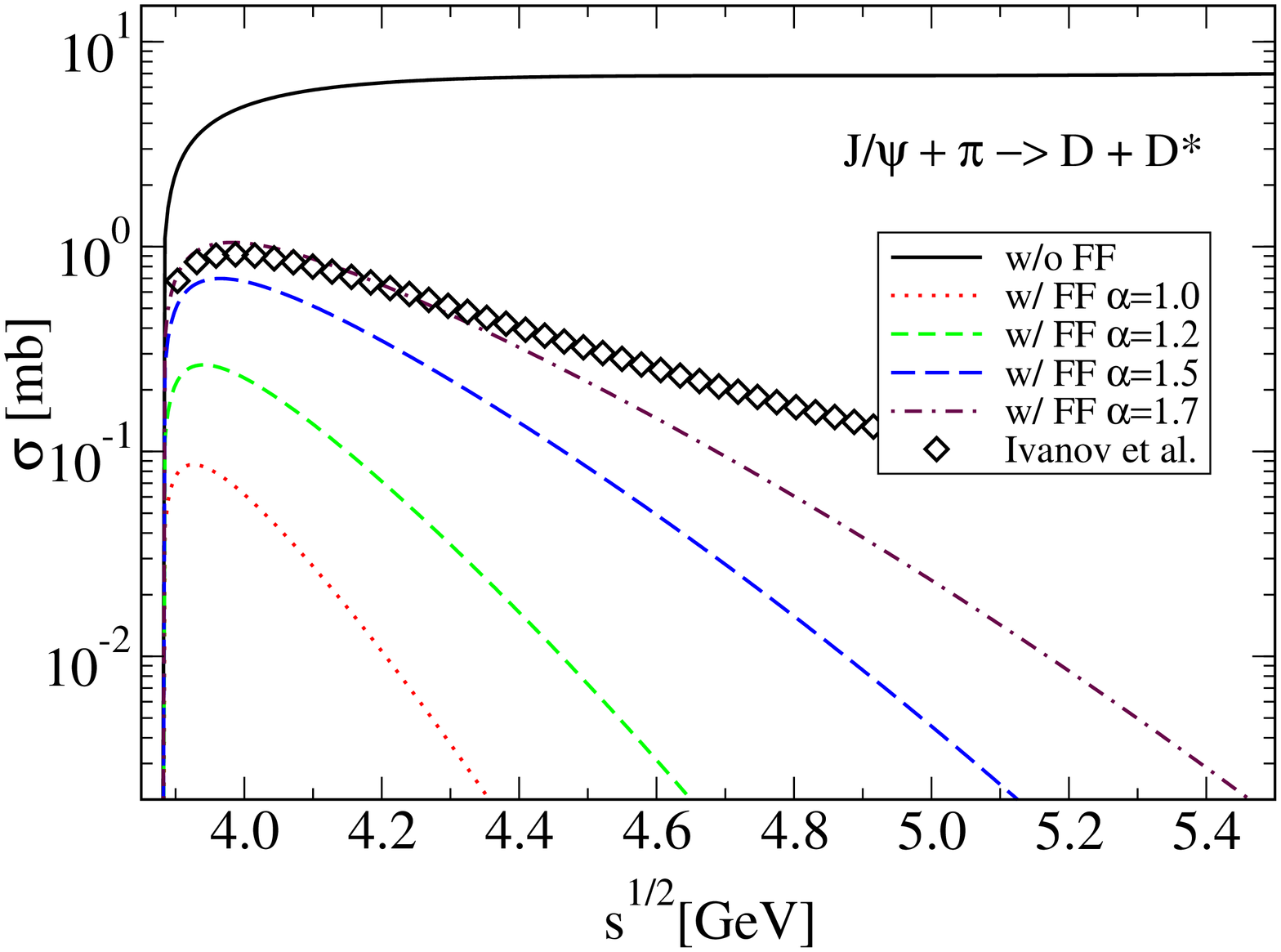,height=0.45\textwidth,angle=-90}
\end{tabular}
\caption{(Color online) Left panel:
$J/\psi$ break-up cross sections in the chiral Lagrangian model with
Gaussian mesonic formfactors and scale factor $\alpha=1$.
Right panel: Dependence of the cross section for the process
$J/\psi+\pi \to D + D^*$ on the scale parameter $\alpha$ in the
mesonic formfactors.
For $\alpha=1.7$ one reproduces the results of the RQM
calculation \cite{Ivanov:2003ge}.
\label{fig3} }
\end{figure}

\begin{figure}[htb]
\epsfig{figure=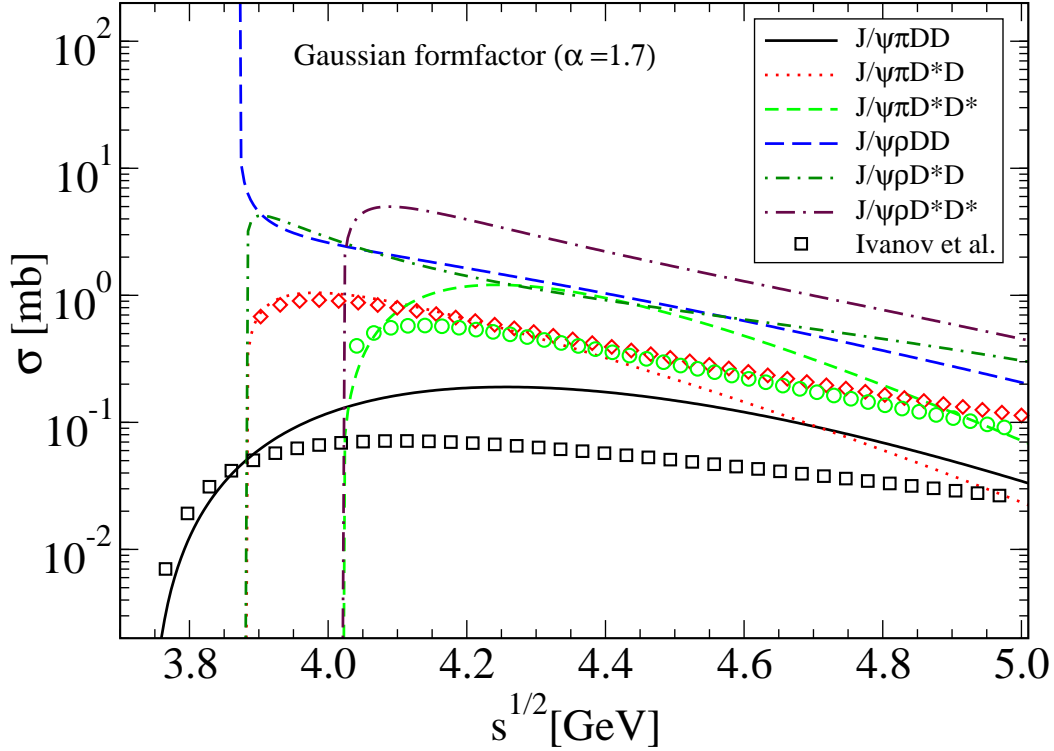,height=0.95\textwidth,angle=-90}
\caption{(Color online) $J/\psi$ break-up cross sections in the chiral
Lagrangian model with Gaussian mesonic formfactors and scale parameter
$\alpha=1.7$. Results from the RQM by Ivanov et al.
\cite{Ivanov:2003ge} are shown for comparison (symbols).
\label{fig4} }
\end{figure}

The results are depicted in Figs. \ref{fig3} and \ref{fig4}.
In the last Section, we discuss the results and their possible
implications for phenomenological applications.

\section{Results}

The $J/\psi$ breakup cross sections by $\pi$ and $\rho$ meson impact have
been formulated within a chiral $SU(4)$ Lagrangian approach including anomalous
processes.
The use of formfactors at the
meson-meson vertices is mandatory since otherwise the high-energy
asymptotics of the processes with hadronic final states will be
overestimated as shown in Fig.~\ref{fig1}.
From a comparison of the results for  the global
formfactor ansatz with a Gaussian function using range parameters
$\Lambda=1$  GeV and $\Lambda=2$ GeV in Fig.~\ref{fig2} we show that
the difference in the
corresponding cross sections above the threshold in the sensitive region
of the energies $\sqrt{s} \simeq 4.5$ is about one order of magnitude.
This underlines the necessity to improve the hadronic formfactor ansatz and
to device a method of the calibration of the range parameters.

In the left panel Fig.~\ref{fig3} we show the energy dependent cross sections 
for different processes using the mesonic formfactor model suggested in
\cite{Ivanov:2001th}, when the range parameters are chosen as in 
Table~\ref{tab:mass}.
In the right panel of the same Figure we vary the parameter $\alpha$ from
$1.0$ to $1.7$ in order calibrate this mesonic formfactor model by comparison
with the result of Ref.~\cite{Ivanov:2003ge} for the  process
$J/\psi + \pi \rightarrow D + \bar{D}*$
dominanting at threshold.

Our calculations show that the guess for the range parameters $\Lambda_i$
successfully reproduces the model calculations of Ref.~\cite{Ivanov:2003ge}
in the energy range up to $\sqrt{s} \simeq 4.4$ GeV for values of
$\alpha  \simeq 1.7 - 2.2$, depending on the process considered.

After fixing the parameter $\alpha = 1.7$ we recalculate the cross sections
of all other processes and show the results in Fig.~\ref{fig4}.
One observes that our calculation gives less suppression than the RQM
calculation.
These calculations allow us to predict dissociation cross sections also for
the $J/\psi$ dissociation by $\rho$-meson impact which are not available in 
the RQM approach by Ivanov et al. \cite{Ivanov:2003ge}.
In comparison to the pion impact processes the $\rho$ meson processes
dominate by a factor $5-8$, basically due to the absence or reduction of the
reaction threshold.


In order to facilitate phenomenological applications of the energy dependent
$J/\psi$ breakup cross sections obtained in this work, we provide a fit in
the form suggested by Barnes et al. in Ref.~\cite{Barnes:2003dg}, 
\begin{equation}
\label{xsfit}
\sigma(s) = \sigma_{\rm max} \left( \varepsilon\over
\varepsilon_{\rm max}\right)^{p_1}
\exp\left(p_2(1-\varepsilon/\varepsilon_{\rm max})\right)~.
\end{equation}
Here $\varepsilon = \sqrt{s} - 2M$ denotes the excitation energy above the
threshold, where $M=(m_3+m_4)/2$ is the mean value of the final state 
$D$-meson masses. 
In Table~\ref{tab:fit} we present the parameter sets obtained from a fit to 
the $J/\psi$ breakup cross sections in the MFCL scheme given in 
Fig.~\ref{fig4}. 
For comparison, the fit parameters corresponding to the RQM results
by Ivanov et al. \cite{Ivanov:2003ge} for pion induced reactions are given in 
brackets.
The value  $\varepsilon_{\rm max}$ corresponds to the excitation energy at 
which the maximum $\sigma_{\rm max}$ of the cross section occurs 
(if it exists).
The parameters $p_1$ and $p_2$ characterize the slopes of the rise above 
threshold and the exponential decay after the maximum, respectively.
The latter is a consequence of the fact that at increasing c.m.s. energy
the overlap between meson wave functions in momentum space decreases, which 
determines the amplitude of the quark exchange process (box diagram in the RQM)
dominating the result for the cross section. 
See also Ref.~\cite{Ivanov:2001th} for a discussion of this point. 
  
\begin{table}
\begin{ruledtabular}
\begin{tabular}{llllll}
process &$\sigma_{\rm max}$[mb]&$M$[GeV]&$\epsilon_{\rm max}$[GeV]&$p_1$&$p_2$ 
\\
\hline
$J/\psi \pi D D $& 0.1912 & 1.824 & 0.6074 & 3.982 & 3.982 \\
& (0.07108)& (1.871) & (0.3741) & (1.024)  & (1.024) \\
$J/\psi \pi D^*D $ &1.048 & 1.940 & 0.1035 & 0.4925 & 0.4925  \\
 &(0.9105) & (1.937) & (0.1198) & (0.4017) & (0.4017) \\
$J/\psi \pi D^*D^* $ &1.215 & 2.002 & 0.2374 & 1.582& 1.582  \\
 &(0.5695) & (2.008) & (0.1332) & (0.4774)&(0.4774)  \\
$J/\psi \rho DD $ &7.156 & 1.939 & 0.00217 & -1.804 & 0.004245 \\
$J/\psi \rho D^*D$ &4.234 & 1.582 & 0.000162 & 0.0004756  & 0.0004756 \\
$J/\psi \rho D^*D^*$ &5.068 & 1.940 & 0.01891 & 0.05066& 0.05066  \\
\end{tabular}
\end{ruledtabular}
\caption{Parameters of the cross section fit (\ref{xsfit}) applied to the 
results of the MFCL model and those of the RQM by Ivanov et al. 
\cite{Ivanov:2003ge} (in brackets), corresponding to Fig.~\ref{fig4}.}
\label{tab:fit}
\end{table}

\section{Conclusion}

The MFCL scheme developed in the present work removes the ambiguity in the
fixation of formfactor parameters by comparison with the RQM approach of
Ref.~\cite{Ivanov:2003ge} and provides a basis for predicting further
$J/\psi$ absorption cross sections.
The first example considered in the present work concerns cross sections for 
breakup processes by $\rho$-meson impact which turn out to be enhanced by one 
order of magnitude over those resulting from pion impact.
This prediction is nicely confirmed by a recent derivation of these cross 
sections within an extended nonlocal RQM \cite{Bourque:2008es}.
A future application of the MFCL approach can consider, e.g., $J/\psi$ 
dissociation by nucleon impact on the basis of a corresponding chiral 
Lagrangian calculation  \cite{Liu:2001yx}.
The result of such a work would provide an essential ingredient for the
study of $J/\psi$ suppression in dense nuclear matter as well as for the
further analysis of cold nuclear matter effects on $J/\psi$ production in 
nuclear collision experiments.


\subsection*{Acknowledgement}
Yu.K. acknowledges support from the Deutsche
Forschungsgemeinschaft under grant no. BL 324/3-1, and RFFI no.
06-01-00228.
The work of D.B. was supported by the Polish Ministry of
Science and Higher Education under grant No. N N 202 0953 33.
H.G. was supported in part by the Heisenberg-Landau program
of the German Ministry for Education and Research (BMBF).


\newpage

\begin{appendix}
\section{Diagrams, amplitudes and couplings}
\label{amplitudes}
\subsection{The process:
$J/\psi (p_1,\mu) + \pi (p_2)\longrightarrow D (p_3) + \bar{D}(p_4)$}
\begin{tabular}{|c|c|c|}
\hline
Diagram & Amplitude & Coupling\\[0cm]
\hline
\begin{minipage}{3cm}
\begin{picture}(80,80)(-20,-20)
 \ArrowLine(0,0)(20,20)
 \ArrowLine(20,20)(0,40)
 \ArrowLine(20,20)(40,0)
 \ArrowLine(40,40)(20,20)
\Text(0,-10)[]{$p_1,\mu$}
\Text(0,50)[]{$p_3$}
\Text(40,50)[]{$p_2$}
\Text(40,-10)[]{$p_4$}
\end{picture}
\end{minipage}
&
$M_\mu^{(1)}=A_1 \epsilon_{\mu\nu\alpha\beta} p_2^\nu p_3^\alpha p_4^\beta$
&
$A_1 = g_{J/\psi \pi D\bar{D}}$
\\[0cm]
\hline
\begin{minipage}{3cm}
\begin{picture}(100,100)(-20,-20)
 \ArrowLine(0,0)(20,20)
 \ArrowLine(20,40)(0,60)
 \ArrowLine(20,20)(40,0)
 \ArrowLine(40,60)(20,40)
\ArrowLine(20,20)(20,40)
\Text(0,-10)[]{$p_1,\mu$}
\Text(0,70)[]{$p_3$}
\Text(40,70)[]{$p_2$}
\Text(40,-10)[]{$p_4$}
\Text(10,30)[]{$Q$}
\Text(30,30)[]{$D^*$}
\end{picture}
\end{minipage}
&
$M_\mu^{(2)}=A_2 \epsilon_{\mu\nu\alpha\beta} p_1^\nu p_4^\alpha
 (p_2)_\sigma N^{\sigma\beta}_{D^*}(Q)$
&$A_2 = {\displaystyle -\frac{g_{J/\psi D^*\bar{D}}~g_{D^*D\pi}}{u-m_{D^*}^2}}$
\\[0cm]
\hline
\begin{minipage}{3cm}
\begin{picture}(80,80)(-10,-10)
 \ArrowLine(0,0)(20,20)
 \ArrowLine(20,20)(0,40)
 \ArrowLine(40,20)(60,0)
 \ArrowLine(60,40)(40,20)
 \ArrowLine(20,20)(40,20)
\Text(0,-10)[]{$p_1,\mu$}
\Text(0,50)[]{$p_3$}
\Text(60,50)[]{$p_2$}
\Text(60,-10)[]{$p_4$}
\Text(30,10)[]{$D^*$}
\Text(30,30)[]{$Q$}
\end{picture}
\vspace*{0.5cm}
\end{minipage}
&$ M_\mu^{(3)}=A_3 \epsilon_{\mu\nu\alpha\beta} p_1^\nu p_3^\alpha (p_2)_\sigma
N^{\sigma\beta}_{D^*}(Q)$
&
$A_3 ={\displaystyle -\frac{g_{J/\psi D^*\bar{D}}~g_{D^*D\pi}}{t-m_{D^*}^2}}$\\
\hline
\end{tabular}

\subsection{The processes:
$J/\psi (p_1,\mu)+ \pi(p_2) \longrightarrow D^*(p_3,\nu) + \bar{D}(p_4),
D(p_3)+\bar{D}^*(p_4,\nu)$ }
\begin{tabular}{|c|c|c|}
\hline
Diagram & Amplitude & Coupling\\[0cm]
\hline
\begin{minipage}{3cm}
\begin{picture}(80,80)(-20,-20)
 \ArrowLine(0,0)(20,20)
 \ArrowLine(20,20)(0,40)
 \ArrowLine(20,20)(40,0)
 \ArrowLine(40,40)(20,20)
\Text(0,-10)[]{$p_1,\mu$}
\Text(0,50)[]{$p_3,\nu$}
\Text(40,50)[]{$p_2$}
\Text(40,-10)[]{$p_4$}
\end{picture}
\end{minipage}
&
$M_{\mu\nu}^{(1)}=C_1 g_{\mu\nu}$
&
$C_1=g_{J/\psi \pi D\bar{D}^*}$
\\[0cm]
\hline
\begin{minipage}{3cm}
\begin{picture}(100,100)(-20,-20)
 \ArrowLine(0,0)(20,20)
 \ArrowLine(20,40)(0,60)
 \ArrowLine(20,20)(40,0)
 \ArrowLine(40,60)(20,40)
\ArrowLine(20,20)(20,40)
\Text(0,-10)[]{$p_1,\mu$}
\Text(0,70)[]{$p_3,\nu$}
\Text(40,70)[]{$p_2$}
\Text(40,-10)[]{$p_4$}
\Text(10,30)[]{$Q$}
\Text(30,30)[]{$D^*$}
\end{picture}
\end{minipage}
&
$
M_{\mu\nu}^{(2)}=C_2 \epsilon_{\mu\gamma\delta\beta}
\epsilon_{\nu\lambda\rho\alpha} p_1^\gamma p_4^\delta p_3^\lambda p_2^\rho
N^{\alpha\beta}_{D^*}(Q)$
&
$C_2={\displaystyle- \frac{g_{J/\psi D^*\bar{D}}~ g_{D^*D^*\pi}}{u-m_{D^*}^2}}$
\\[0cm]
\hline
\begin{minipage}{3cm}
\begin{picture}(80,80)(-20,-20)
 \ArrowLine(0,0)(20,20)
 \ArrowLine(20,20)(0,40)
 \ArrowLine(40,20)(60,0)
 \ArrowLine(60,40)(40,20)
 \ArrowLine(20,20)(40,20)
\Text(0,-10)[]{$p_1,\mu$}
\Text(0,50)[]{$p_3,\nu$}
\Text(60,50)[]{$p_2$}
\Text(60,-10)[]{$p_4$}
\Text(30,10)[]{$D^*$}
\Text(30,30)[]{$Q$}
\end{picture}
\end{minipage}
&
\begin{minipage}{10cm}
\begin{eqnarray}
M_{\mu\nu}^{(3)}&=&C_3 \left[2g_{\alpha\nu}(p_3)_\mu-
g_{\mu\nu}(p_1+p_3)_\alpha+2g_{\alpha\mu}(p_1)_\nu
\right] \nonumber \\
&&N^{\alpha\beta}_{D^*}(Q)(p_2+p_4)_\beta\nonumber
\end{eqnarray}
\end{minipage}
&
$C_3 = {\displaystyle\frac{g_{J/\psi D^*\bar{D}^*}~g_{D^*D\pi}}{t-m_{D^*}^2}}$
\\[0cm]
\hline
\begin{minipage}{3cm}
\begin{picture}(100,100)(-20,-20)
 \ArrowLine(0,0)(20,20)
 \ArrowLine(20,40)(0,60)
 \ArrowLine(20,20)(40,0)
 \ArrowLine(40,60)(20,40)
\ArrowLine(20,20)(20,40)
\Text(0,-10)[]{$p_1,\mu$}
\Text(0,70)[]{$p_3,\nu$}
\Text(40,70)[]{$p_2$}
\Text(40,-10)[]{$p_4$}
\Text(10,30)[]{$Q$}
\Text(30,30)[]{$D$}
\end{picture}
\end{minipage}
&
$M_{\mu\nu}^{(4)}=C_4 ~4~(p_2)_\nu (p_4)_\mu $
&
$C_4 = {\displaystyle - \frac{g_{J/\psi D\bar{D}}~ g_{D^*D\pi}}{u-m_{D}^2}} $
\\[0cm]
\hline
\end{tabular}

\subsection{The process: $J/\psi(p_1,\mu) + \pi(p_2)
\longrightarrow D^* (p_3,\nu)+ \bar{D}^*(p_4,\lambda)$ }
\begin{tabular}{|c|c|c|}
\hline
Diagram & Amplitude & Coupling\\[0cm]
\hline
\begin{minipage}{3cm}
\begin{picture}(80,80)(-20,-20)
 \ArrowLine(0,0)(20,20)
 \ArrowLine(20,20)(0,40)
 \ArrowLine(20,20)(40,0)
 \ArrowLine(40,40)(20,20)
\Text(0,-10)[]{$p_1,\mu$}
\Text(0,50)[]{$p_3,\nu$}
\Text(40,50)[]{$p_2$}
\Text(40,-10)[]{$p_4,\lambda$}
\end{picture}
\end{minipage}
&
$\displaystyle M_{\mu\nu\lambda}^{(1)}=B_{11} \epsilon_{\mu\nu\lambda\beta }p_2^\beta
+B_{12}\epsilon_{\mu\nu\lambda\beta}p_1^\beta$
&
\begin{minipage}{3cm}
\begin{eqnarray}
B_{11} & = & g_{J/\psi \pi D^*\bar{D^*}}\nonumber \\
B_{12} & = & g_{J/\psi \pi D^*\bar{D^*}}\nonumber
\end{eqnarray}
\end{minipage}
\\[0cm]
\hline
\begin{minipage}{3cm}
\begin{picture}(100,100)(-20,-20)
 \ArrowLine(0,0)(20,20)
 \ArrowLine(20,40)(0,60)
 \ArrowLine(20,20)(40,0)
 \ArrowLine(40,60)(20,40)
\ArrowLine(20,20)(20,40)
\Text(0,-10)[]{$p_1,\mu$}
\Text(0,70)[]{$p_3,\nu$}
\Text(40,70)[]{$p_2$}
\Text(40,-10)[]{$p_4,\lambda$}
\Text(10,30)[]{$Q$}
\Text(30,30)[]{$D^*$}
\end{picture}
\end{minipage}
&
\begin{minipage}{10cm}
\begin{eqnarray}
 M_{\mu\nu\lambda}^{(2)}&=&B_2 \epsilon_{\nu\gamma\delta\alpha}\left[ 2 g_{\mu\beta} (p_1)_\lambda-
g_{\mu\lambda}(p_1+p_4)_\beta+2g_{\lambda\beta}(p_4)_\mu \right]\nonumber\\
&& N^{\alpha\beta}_{D^*}(Q)p_3^\gamma
p_2^\delta
\nonumber
\end{eqnarray}
\end{minipage}
&
$B_2={\displaystyle \frac{g_{J/\psi D^*\bar{D^*}}~g_{D^*D^*\pi}}{u-m_{D^*}^2}}$
\\[0cm]
\hline
\begin{minipage}{3cm}
\begin{picture}(80,80)(-15,-20)
 \ArrowLine(0,0)(20,20)
 \ArrowLine(20,20)(0,40)
 \ArrowLine(40,20)(60,0)
 \ArrowLine(60,40)(40,20)
 \ArrowLine(20,20)(40,20)
\Text(0,-10)[]{$p_1,\mu$}
\Text(0,50)[]{$p_3,\nu$}
\Text(60,50)[]{$p_2$}
\Text(60,-10)[]{$p_4,\lambda$}
\Text(30,10)[]{$D^*$}
\Text(30,30)[]{$Q$}
\end{picture}
\end{minipage}
&
\begin{minipage}{10cm}
\begin{eqnarray}
 M_{\mu\nu\lambda}^{(3)}&=&B_3 \epsilon_{\lambda\gamma\delta\alpha}\left[ 2 g_{\nu\beta} (p_3)_\mu-
g_{\mu\nu}(p_1+p_3)_\beta+2g_{\mu\beta}(p_1)_\nu \right]\nonumber\\
&&N^{\alpha\beta}_{D^*}(Q)p_2^\gamma p_4^\delta
\nonumber
\end{eqnarray}
\end{minipage}
&
$B_3={\displaystyle \frac{g_{J/\psi D^*\bar{D^*}}~g_{D^*D^*\pi}}{t-m_{D^*}^2}}$
\\[0cm]
\hline
\begin{minipage}{3cm}
\begin{picture}(100,100)(-20,-20)
 \ArrowLine(0,0)(20,20)
 \ArrowLine(20,40)(0,60)
 \ArrowLine(20,20)(40,0)
 \ArrowLine(40,60)(20,40)
\ArrowLine(20,20)(20,40)
\Text(0,-10)[]{$p_1,\mu$}
\Text(0,70)[]{$p_3,\nu$}
\Text(40,70)[]{$p_2$}
\Text(40,-10)[]{$p_4,\lambda$}
\Text(10,30)[]{$Q$}
\Text(30,30)[]{$D$}
\end{picture}
\end{minipage}
&
$M_{\mu\nu\lambda}^{(4)}=B_4  \epsilon_{\mu\lambda\alpha\beta}
p_1^\alpha p_4^\beta (p_2)_\nu$
&
$B_4 = {\displaystyle -\frac{2g_{J/\psi D^*\bar{D}}~g_{D^*D\pi}}{u-m_{D}^2}}$
\\[0cm]
\hline
\begin{minipage}{3cm}
\begin{picture}(80,80)(-15,-20)
 \ArrowLine(0,0)(20,20)
 \ArrowLine(20,20)(0,40)
 \ArrowLine(40,20)(60,0)
 \ArrowLine(60,40)(40,20)
 \ArrowLine(20,20)(40,20)
\Text(0,-10)[]{$p_1,\mu$}
\Text(0,50)[]{$p_3,\nu$}
\Text(60,50)[]{$p_2$}
\Text(60,-10)[]{$p_4,\lambda$}
\Text(30,10)[]{$D$}
\Text(30,30)[]{$Q$}
\end{picture}
\end{minipage}
&
$M_{\mu\nu\lambda}^{(5)}=B_5  \epsilon_{\mu\nu\alpha\beta}
p_1^\alpha p_3^\beta (p_2)_\lambda $
&
$B_5= {\displaystyle \frac{2g_{J/\psi D^*\bar{D}}~g_{D^*D\pi}}{t-m_{D}^2} } $
\\[0cm]
\hline
\end{tabular}

\subsection{The process:
$J/\psi (p_{1,\mu})+ \rho(p_{2,\nu}) \longrightarrow D(p_3) + \bar{D}(p_4)$}
\begin{tabular}{|c|c|c|}
\hline
Diagram & Amplitude & Coupling\\[0cm]
\hline
\begin{minipage}{3cm}
\begin{picture}(80,80)(-20,-20)
 \ArrowLine(0,0)(20,20)
 \ArrowLine(20,20)(0,40)
 \ArrowLine(20,20)(40,0)
 \ArrowLine(40,40)(20,20)
\Text(0,-10)[]{$p_1,\mu$}
\Text(0,50)[]{$p_3$}
\Text(40,50)[]{$p_2,\nu$}
\Text(40,-10)[]{$p_4$}
\Text(120,20)[]{}
\end{picture}
\end{minipage}
&
$M_{\mu\nu}^{(1)}= G_1 \delta_{\mu\nu}$
&
$G_1 = g_{J/\psi \rho D\bar{D}}$
\\[0cm]
\hline
\begin{minipage}{3cm}
\begin{picture}(100,100)(-20,-20)
 \ArrowLine(0,0)(20,20)
 \ArrowLine(20,40)(0,60)
 \ArrowLine(20,20)(40,0)
 \ArrowLine(40,60)(20,40)
\ArrowLine(20,20)(20,40)
\Text(0,-10)[]{$p_1,\mu$}
\Text(0,70)[]{$p_3$}
\Text(40,70)[]{$p_2,\nu$}
\Text(40,-10)[]{$p_4$}
\Text(10,30)[]{$Q$}
\Text(30,30)[]{$D$}
\end{picture}
\end{minipage}
&
$M_{\mu\nu}^{(2)}=G_2 4 (p_3)_\nu (p_4)_\mu$
&
$G_2= {\displaystyle \frac{g_{J/\psi D\bar{D}}~g_{DD\rho}}{u-m_{D}^2}}  $
\\[0cm]
\hline
\begin{minipage}{3cm}
\begin{picture}(80,80)(-15,-20)
 \ArrowLine(0,0)(20,20)
 \ArrowLine(20,20)(0,40)
 \ArrowLine(40,20)(60,0)
 \ArrowLine(60,40)(40,20)
 \ArrowLine(20,20)(40,20)
\Text(0,-10)[]{$p_1,\mu$}
\Text(0,50)[]{$p_3$}
\Text(60,50)[]{$p_2,\nu$}
\Text(60,-10)[]{$p_4$}
\Text(30,10)[]{$D$}
\Text(30,30)[]{$Q$}
\end{picture}
\end{minipage}
&
$M_{\mu\nu}^{(3)}=G_3 4 (p_3)_\mu (p_4)_\nu$
&
$G_3= {\displaystyle \frac{g_{J/\psi D\bar{D} }~g_{DD\rho}}{t-m_{D}^2}}$
\\[0cm]
\hline
\begin{minipage}{3cm}
\begin{picture}(100,100)(-20,-20)
 \ArrowLine(0,0)(20,20)
 \ArrowLine(20,40)(0,60)
 \ArrowLine(20,20)(40,0)
 \ArrowLine(40,60)(20,40)
\ArrowLine(20,20)(20,40)
\Text(0,-10)[]{$p_1,\mu$}
\Text(0,70)[]{$p_3$}
\Text(40,70)[]{$p_2,\nu$}
\Text(40,-10)[]{$p_4$}
\Text(10,30)[]{$Q$}
\Text(30,30)[]{$D^*$}
\end{picture}
\end{minipage}
&
$\displaystyle M_{\mu\nu}^{(4)}=G_4
\epsilon_{\nu\rho\sigma\alpha}\epsilon_{\mu\lambda\gamma\beta}p_1^{\lambda}p_2^{\sigma}p_3^{\rho}p_4^{\gamma}
N^{\alpha\beta}_{D^*}(Q)
$
&
$G_4 = {\displaystyle \frac{g_{J/\psi D\bar{D^*}}~g_{D*D\rho}}{u-m_{D^*}^2}} $
\\[0cm]
\hline
\begin{minipage}{3cm}
\begin{picture}(80,80)(-15,-20)
 \ArrowLine(0,0)(20,20)
 \ArrowLine(20,20)(0,40)
 \ArrowLine(40,20)(60,0)
 \ArrowLine(60,40)(40,20)
 \ArrowLine(20,20)(40,20)
\Text(0,-10)[]{$p_1,\mu$}
\Text(0,50)[]{$p_3$}
\Text(60,50)[]{$p_2,\nu$}
\Text(60,-10)[]{$p_4$}
\Text(30,10)[]{$D^*$}
\Text(30,30)[]{$Q$}
\end{picture}
\end{minipage}
&
$\displaystyle M_{\mu\nu}^{(5)}=G_5
\epsilon_{\nu\rho\sigma\alpha}\epsilon_{\mu\lambda\gamma\beta}p_1^{\lambda}p_2^{\sigma}p_3^{\gamma}p_4^{\rho}
N^{\alpha\beta}_{D^*}(Q)$
&
$G_5 = {\displaystyle \frac{g_{J/\psi D\bar{D^*}}~g_{D^*D\rho}}{t-m_{D^*}^2}}$
\\[0cm]
\hline
\end{tabular}

\subsection{The process:
$J/\psi (p_1,\mu)+ \rho(p_2,\nu) \longrightarrow D^* (p_3,\lambda)+
 \bar{D}(p_4), D(p_3)+\bar{D}^* (p_4,\lambda)$}
\begin{tabular}{|c|c|c|}
\hline
Diagram & Amplitude & Coupling\\[0cm]
\hline
\begin{minipage}{3cm}
\begin{picture}(80,80)(-20,-20)
 \ArrowLine(0,0)(20,20)
 \ArrowLine(20,20)(0,40)
 \ArrowLine(20,20)(40,0)
 \ArrowLine(40,40)(20,20)
\Text(0,-10)[]{$p_1,\mu$}
\Text(0,50)[]{$p_3,\lambda$}
\Text(40,50)[]{$p_2,\nu$}
\Text(40,-10)[]{$p_4$}
\end{picture}
\end{minipage}
&
$M_{\mu\nu\lambda}^{(1)}=
H_{11} \epsilon_{\mu\nu\lambda\beta}p_3^\beta +
H_{12}  \epsilon_{\mu\nu\lambda\beta}p_4^\beta $
&
\begin{minipage}{3cm}
\begin{eqnarray}
H_{11} & = &  g_{J/\psi \rho D^*D}\nonumber  \\
H_{12}& = &  g_{J/\psi  \rho DD^*} \nonumber
\end{eqnarray}
\end{minipage}
\\[0cm]
\hline
\begin{minipage}{3cm}
\begin{picture}(100,100)(-20,-20)
 \ArrowLine(0,0)(20,20)
 \ArrowLine(20,40)(0,60)
 \ArrowLine(20,20)(40,0)
 \ArrowLine(40,60)(20,40)
\ArrowLine(20,20)(20,40)
\Text(0,-10)[]{$p_1,\mu$}
\Text(0,70)[]{$p_3$}
\Text(40,70)[]{$p_2,\nu$}
\Text(40,-10)[]{$p_4$}
\Text(10,30)[]{$Q$}
\Text(30,30)[]{$D^*$}
\end{picture}
\end{minipage}
&
\begin{minipage}{10cm}
\begin{eqnarray}
M_{\mu\nu\lambda}^{(2)}&=&H_2
\epsilon_{\mu\rho\sigma\alpha}
\left[ 2g_{\lambda\beta}(p_3)_\nu-g_{\lambda\nu}(p_2+p_3)_\beta
+2(p_2)_\lambda g_{\nu\beta}\right]
\nonumber \\
&&N^{\alpha\beta}_{D^*}(Q)p_1^{\rho}p_4^{\sigma}
\nonumber
\end{eqnarray}
\end{minipage}
&
$H_2={\displaystyle \frac{g_{\rho D^*\bar{D}^*}~g_{J/\psi D^*D}}{u-m_{D^*}^2}}$
\\[0cm]
\hline
\begin{minipage}{3cm}
\begin{picture}(80,80)(-15,-20)
 \ArrowLine(0,0)(20,20)
 \ArrowLine(20,20)(0,40)
 \ArrowLine(40,20)(60,0)
 \ArrowLine(60,40)(40,20)
 \ArrowLine(20,20)(40,20)
\Text(0,-10)[]{$p_1,\mu$}
\Text(0,50)[]{$p_3,\lambda$}
\Text(60,50)[]{$p_2,\nu$}
\Text(60,-10)[]{$p_4$}
\Text(30,10)[]{$D^*$}
\Text(30,30)[]{$Q$}
\end{picture}
\end{minipage}
&
\begin{minipage}{10cm}
\begin{eqnarray}
M_{\mu\nu\lambda}^{(3)}&=&H_3
\epsilon_{\nu\rho\sigma\alpha}
\left[ 2g_{\lambda\beta}(p_3)_\mu-g_{\lambda\mu}(p_1+p_3)_\beta
+2 g_{\mu\beta}(p_1)_\lambda\right]
\nonumber\\
&& N^{\alpha\beta}_{D^*}(Q)p_2^{\rho}p_4^{\sigma}
\nonumber
\end{eqnarray}
\end{minipage}
&
$H_3 ={\displaystyle -\frac{g_{J/\psi D^*\bar{D^*}}~g_{\rho D^*D}}{t-m_{D^*}^2} }$
\\[0cm]
\hline
\begin{minipage}{3cm}
\begin{picture}(100,100)(-20,-20)
 \ArrowLine(0,0)(20,20)
 \ArrowLine(20,40)(0,60)
 \ArrowLine(20,20)(40,0)
 \ArrowLine(40,60)(20,40)
\ArrowLine(20,20)(20,40)
\Text(0,-10)[]{$p_1,\mu$}
\Text(0,70)[]{$p_3,\lambda$}
\Text(40,70)[]{$p_2,\nu$}
\Text(40,-10)[]{$p_4$}
\Text(10,30)[]{$Q$}
\Text(30,30)[]{$D$}
\end{picture}
\end{minipage}
&
$M_{\mu\nu\lambda}^{(4)}=H_4 \epsilon_{\nu\lambda\alpha\beta} p_2^\alpha p_3^\beta (p_4)_\mu$
&
$H_4= {\displaystyle - \frac{2g_{J/\psi D\bar{D}}~g_{\rho D^*D}}{u-m_{D}^2} } $
\\[0cm]
\hline
\begin{minipage}{3cm}
\begin{picture}(80,80)(-15,-20)
 \ArrowLine(0,0)(20,20)
 \ArrowLine(20,20)(0,40)
 \ArrowLine(40,20)(60,0)
 \ArrowLine(60,40)(40,20)
 \ArrowLine(20,20)(40,20)
\Text(0,-10)[]{$p_1,\mu$}
\Text(0,50)[]{$p_3,\lambda$}
\Text(60,50)[]{$p_2,\nu$}
\Text(60,-10)[]{$p_4$}
\Text(30,10)[]{$D$}
\Text(30,30)[]{$Q$}
\end{picture}
\end{minipage}
&
$M_{\mu\nu\lambda}^{(5)}=H_5 \epsilon_{\mu\lambda\alpha\beta} p_1^\alpha p_3^\beta (p_4)_\nu$
&
$H_5= {\displaystyle \frac{2 g_{J/\psi D^*\bar{D}}~g_{\rho D D}}{t-m_{D}^2}} $
\\[0cm]
\hline
\end{tabular}

\subsection{The processes: $J/\psi (p_1,\mu)+ \rho (p_2,\nu)
\longrightarrow D^*(p_3,\lambda) + \bar{D}^*(p_4,\rho)$}
\begin{tabular}{|c|c|c|}
\hline
Diagram & Amplitude & Coupling\\[0cm]
\hline
\begin{minipage}{3cm}
\begin{picture}(80,80)(-20,-20)
 \ArrowLine(0,0)(20,20)
 \ArrowLine(20,20)(0,40)
 \ArrowLine(20,20)(40,0)
 \ArrowLine(40,40)(20,20)
\Text(0,-10)[]{$p_1,\mu$}
\Text(0,50)[]{$p_3,\lambda$}
\Text(40,50)[]{$p_2,\nu$}
\Text(40,-10)[]{$p_4,\rho$}
\end{picture}
\end{minipage}
&
$M_{\mu\nu\lambda\rho}^{(1)}=
K_1 \left[2 g_{\mu\nu}g_{\rho\lambda}-
g_{\mu\rho}g_{\nu\lambda}-g_{\mu\lambda}g_{\nu\rho}\right]$
&
$K_{1} = g_{J/\psi \rho D^*\bar{D^*}}$
\\[0cm]
\hline
\begin{minipage}{3cm}
\begin{picture}(100,100)(-20,-20)
 \ArrowLine(0,0)(20,20)
 \ArrowLine(20,40)(0,60)
 \ArrowLine(20,20)(40,0)
 \ArrowLine(40,60)(20,40)
\ArrowLine(20,20)(20,40)
\Text(0,-10)[]{$p_1,\mu$}
\Text(0,70)[]{$p_3,\lambda$}
\Text(40,70)[]{$p_2,\nu$}
\Text(40,-10)[]{$p_4,\rho$}
\Text(10,30)[]{$Q$}
\Text(30,30)[]{$D^*$}
\end{picture}
\end{minipage}
&
\begin{minipage}{10cm}
\begin{eqnarray}
M_{\mu\nu\lambda\rho}^{(2)}&=&
K_2 \left[ 2g_{\rho\alpha}(p_4)_\mu -g_{\mu\rho}(p_1+p_4)_\alpha
+2 g_{\mu\alpha}(p_1)_\rho \right]
\nonumber \\
&&N^{\alpha\beta}_{D^*}(Q)
\nonumber \\
&&
 \left[2g_{\beta\lambda}(p_3)_\nu -g_{\nu\lambda} (p_2+p_3)_\beta
+2g_{\beta\nu} (p_2)_\lambda \right]
\nonumber
\end{eqnarray}
\end{minipage}
&
$K_2= {\displaystyle \frac{g_{J/\psi D^*\bar{D^*}}~g_{\rho D^*D^*}}{u-m_{D^*}^2}}$
\\[0cm]
\hline
\begin{minipage}{3cm}
\begin{picture}(80,80)(-15,-20)
 \ArrowLine(0,0)(20,20)
 \ArrowLine(20,20)(0,40)
 \ArrowLine(40,20)(60,0)
 \ArrowLine(60,40)(40,20)
 \ArrowLine(20,20)(40,20)
\Text(0,-10)[]{$p_1,\mu$}
\Text(0,50)[]{$p_3,\lambda$}
\Text(60,50)[]{$p_2,\nu$}
\Text(60,-10)[]{$p_4,\rho$}
\Text(30,10)[]{$D^*$}
\Text(30,30)[]{$Q$}
\end{picture}
\end{minipage}
&
\begin{minipage}{10cm}
\begin{eqnarray}
M_{\mu\nu\lambda\rho}^{(3)}&=&
K_3 \left[ 2g_{\lambda\alpha}(p_3)_\mu -(p_1+p_3)_\alpha
         g_{\mu\lambda}+2g_{\mu\alpha}p_1^\lambda\right]
 \nonumber \\
&&N^{\alpha\beta}_{D^*}(Q)
 \nonumber \\
&&
\left[ 2 g_{\beta\rho}(p_4)_\nu-g_{\nu\rho}(p_2+p_4)_\beta
+2g_{\beta\nu}(p_2)_\rho\right]
\nonumber
\end{eqnarray}
\end{minipage}
&
$K_3= {\displaystyle \frac{g_{J/\psi D^*\bar{D^*}}~g_{\rho D^*D^*}}{t-m_{D^*}^2} }$
\\[0cm]
\hline
\begin{minipage}{3cm}
\begin{picture}(100,100)(-20,-20)
 \ArrowLine(0,0)(20,20)
 \ArrowLine(20,40)(0,60)
 \ArrowLine(20,20)(40,0)
 \ArrowLine(40,60)(20,40)
\ArrowLine(20,20)(20,40)
\Text(0,-10)[]{$p_1,\mu$}
\Text(0,70)[]{$p_3,\lambda$}
\Text(40,70)[]{$p_2,\nu$}
\Text(40,-10)[]{$p_4,\rho$}
\Text(10,30)[]{$Q$}
\Text(30,30)[]{$D$}
\end{picture}
\end{minipage}
&
$M_{\mu\nu\lambda\rho}^{(4)}=K_4  \epsilon_{\mu\rho\alpha\beta}
\epsilon_{\nu\lambda\gamma\delta}
p_1^\alpha p_2^{\gamma}p_3^{\delta} p_4^\beta $
&
$K_4 =  {\displaystyle \frac{g_{J/\psi D^*\bar{D}}~g_{\rho D^*D}}{u-m_{D}^2} }$
\\[0cm]
\hline
\begin{minipage}{3cm}
\begin{picture}(80,80)(-15,-20)
 \ArrowLine(0,0)(20,20)
 \ArrowLine(20,20)(0,40)
 \ArrowLine(40,20)(60,0)
 \ArrowLine(60,40)(40,20)
 \ArrowLine(20,20)(40,20)
\Text(0,-10)[]{$p_1,\mu$}
\Text(0,50)[]{$p_3,\lambda$}
\Text(60,50)[]{$p_2,\nu$}
\Text(60,-10)[]{$p_4,\rho$}
\Text(30,10)[]{$D$}
\Text(30,30)[]{$Q$}
\end{picture}
\end{minipage}
&
$M_{\mu\nu\lambda\rho}^{(5)}=K_5  \epsilon_{\mu\lambda\alpha\beta}
\epsilon_{\nu\rho\gamma\delta}
p_1^\alpha p_2^{\gamma} p_3^\beta p_4^{\delta}$
&
$K_5= {\displaystyle  \frac{g_{J/\psi D^*\bar{D}}~g_{\rho D^*D}}{t-m_{D}^2} } $
\\[0cm]
\hline
\end{tabular}
\section{Coupling constants}
\label{constants}
Here we summarize the values of the coupling constants occuring in the
amplitudes of the Appendix \ref{amplitudes}. They correspond to those given
in Ref.~\cite{Oh:2000qr}

\begin{table}
\label{couplings}
\begin{tabular}{ll}
\hline\hline
Coupling constants&Value
\\
\hline
$g_{J/\psi \pi D {D}}$ &  16.0 ~GeV$^{-3}$\\
$g_{J/\psi \pi D {D^*}}$ &  33.92 \\
$g_{J/\psi \pi D^* {D^*}}$ &  38.19 ~GeV$^{-1}$\\
\hline
$g_{J/\psi \rho D {D}}$ &  38.86 \\
$g_{J/\psi \rho D {D^*}}$ &  21.77~GeV$^{-1}$\\
$g_{J/\psi \rho D^* {D^*}}$ &  19.43 \\
\hline
\hline
$g_{J/\psi D {D}}$  &  7.71\\
$g_{J/\psi D^* {D} }$ & 8.64~GeV$^{-1}$\\
$g_{J/\psi D^* {D^*} }$ &  7.71\\
\hline
$g_{D^*D\pi} $ &  8.84\\
$g_{D^*D^*\pi}$  &   9.08~GeV$^{-1}$\\
\hline
$g_{DD\rho} $ &   2.52\\
$g_{D*D\rho} $ &  2.82~GeV$^{-1}$\\
$g_{D^*D^*\rho} $ & 2.52\\
\hline\hline
\end{tabular}
\caption{Values for the coupling constants introduced in the
amplitudes for the processes given in Appendix \ref{amplitudes}.}
\end{table}

\end{appendix}

\end{document}